\documentclass[conference]{IEEEtran}
\IEEEoverridecommandlockouts

\usepackage{cite}
\usepackage{amsmath,amssymb,amsfonts}
\usepackage{algorithmic}
\usepackage{graphicx}
\usepackage{textcomp}
\usepackage{xcolor}
\def\BibTeX{{\rm B\kern-.05em{\sc i\kern-.025em b}\kern-.08em
    T\kern-.1667em\lower.7ex\hbox{E}\kern-.125emX}}
\begin{document}

\IEEEoverridecommandlockouts
\IEEEpubid{\makebox[\columnwidth]{979-8-3315-0376-5/25/\$31.00 $\copyright$2025 IEEE \hfill}\hspace{\columnsep}\makebox[\columnwidth]{ }}

\title{ReWeave: Traffic Engineering with Robust Path Weaving for Localized Link Failure Recovery\\
}

\author{
\IEEEauthorblockN{Jingyi Guan}
\IEEEauthorblockA{\textit{College of Computer Science} \\ 
\textit{and Artificial Intelligence} \\
\textit{Fudan University} \\
Shanghai, China \\
23301020098@m.fudan.edu.cn}
\and
\IEEEauthorblockN{Kun Qiu}
\IEEEauthorblockA{\textit{Institute of Space Internet} \\
\textit{Fudan University} \\
Shanghai, China \\
qkun@fudan.edu.cn}
\and
\IEEEauthorblockN{Jin Zhao}
\IEEEauthorblockA{\textit{College of Computer Science} \\ 
\textit{and Artificial Intelligence} \\
\textit{Fudan University} \\
Shanghai, China \\
jzhao@fudan.edu.cn}
}

\maketitle

\begin{abstract}
Link failures occur frequently in Internet Service Provider (ISP) networks and pose significant challenges for Traffic Engineering (TE). Existing TE schemes either reroute traffic over vulnerable static paths, leading to performance degradation, or precompute backup routes for a broad range of failure scenarios, which introduces high overhead and limits scalability. Hence, an effective failure recovery mechanism is required to offer sufficient path diversity under constrained overhead, thereby ensuring robust and performant network operation.
This paper presents ReWeave, a scalable and efficient link-level TE scheme that enables localized rerouting by equipping each link with a compact set of adjacent-only backup paths. Upon detecting a failure, only the routers at both ends of the failed link reroute traffic dynamically using SRv6-based detours, without controller intervention or full-path recomputation. 
Evaluation results on large-scale backbone networks demonstrate that ReWeave outperforms existing TE schemes in link failure scenarios. 
Compared to HARP, the state-of-the-art failure recovery scheme based on centralized control and dynamic traffic reallocation, our approach reduces the average maximum link utilization by 10.5\%--20.1\%, and lowers the worst-case utilization by 29.5\%--40.9\%. 
When compared with Flexile, a protection-based scheme that precomputes routes for multi-failure scenarios, ReWeave achieves a similarly low packet loss rate in 90\% of failure cases, while maintaining a response speed comparable to the fastest router-based local rerouting schemes.

\end{abstract}

\begin{IEEEkeywords}
Traffic Engineering, Link Failure Recovery, Load Balancing, Routing Optimization.
\end{IEEEkeywords}

\section{Introduction}
With the expansion of network scale, Internet Service Providers (ISPs) require effective mechanisms to manage traffic across their backbone WANs\cite{b1,b3}. Traffic Engineering (TE), through a centralized controller in Software-Defined Networking (SDN)\cite{b4}, periodically makes routing decisions and deploys them to routers. This enables ISPs to achieve specific network-wide optimization objectives, such as minimizing maximum link utilization\cite{b5}. To ensure uninterrupted data transmission, ISPs aim to maintain satisfactory network performance even in the event of failures\cite{b9}. 
Studies have shown that approximately 70\% of unplanned network failures are caused by single-link failures—a scenario fundamentally distinct from more complex node or correlated failures—making them the most critical challenge for network stability \cite{b6}.
Therefore, TE needs to effectively address link failure scenarios and sustain decision-making performance\cite{b11}. 

Current TE schemes consist of two stages: path selection and proportional allocation\cite{b9}. The former is constrained by switch flow table capacity and configuration latency, making frequent execution infeasible\cite{b12}. As a result, practical deployments typically adjust traffic ratios dynamically over a static set of preselected paths. When a link failure occurs, it fundamentally changes the network topology, requiring the TE system to react quickly to maintain performance.



Existing failure handling mechanisms fall into two categories: protection and restoration \cite{b13}. Protection schemes can incur high overhead and limited scalability, as they must pre-install backup paths for numerous failure scenarios\cite{b10,b14}. Restoration schemes react to failures, but both of its main approaches—local ratio adjustments and centralized recomputation\cite{b9,b15}—are fundamentally constrained by the same structural limitation: they must reallocate traffic over the surviving static paths, which often become fragile and congested after a topology change.



A key challenge in failure recovery is that the effectiveness of ML-based TE critically depends on the underlying path set's stability. Simply deleting failed paths shrinks the set, invalidates pre-trained allocation ratios, and degrades performance. A better approach should therefore preserve the path set by reusing non-failed links. The challenge is bypassing the single failed link. Our key insight is that path sets between adjacent nodes are inherently robust. This motivated ReWeave, a scheme that provisions local backup paths for each link. By patching the failed segment locally, ReWeave repairs the original global path, thus maintaining a robust path set where the pre-trained ML model remains fully effective.

When a router detects a link failure, the routers at both ends of the failed link can quickly reroute the affected traffic onto these alternative paths. Specifically, SDN routers can learn about link failures through implicit probes carried in the data packets and subsequently obtain alternative paths to reach the router at the other end of the failed link. For traffic destined to traverse the failed link, the router can determine the next-hop forwarding destination and, by modifying the Segment Identifier (SID) list of the packets passing through the link, steer the data flow along an alternative path to reach the router at the other end of the failed link without traversing the failed link. This approach reduces the number of variables that need to be considered for rerouting from $O(n^2)$ to $O(e)$, significantly lowering the overhead and keeping it within an acceptable range. This localized mechanism preserves the size of the global path set, allowing the pre-trained TE model to remain fully effective without costly retraining. While maintaining a packet loss rate comparable to protection-based schemes\cite{b14}, ReWeave achieves a reaction speed on par with router-based rerouting schemes, and reduces the average Maximum Link Utilization (MLU) by 4.1\% to 20.2\%, with the worst-case reduction reaching 37.8\%. Compared to the state-of-the-art failure recovery scheme HARP \cite{b15}, ReWeave reduces the average MLU by 10.5\% to 20.1\%, and achieves a maximum reduction of 40.9\% in the worst case.

The contributions of this paper are summarized as follows:
\begin{itemize}
    \item We propose a novel localized failure recovery paradigm that reformulates the problem to focus on link-level impact. We design and implement this in ReWeave, a framework that uses pre-installed, adjacent-node backup paths and SRv6 for fast, controller-free recovery.
    \item We demonstrate how this localized mechanism uniquely synergizes with learning-based TE, preserving the integrity of the path set to maintain consistent traffic allocation and avoid costly model retraining under failures.
    \item We conduct comprehensive evaluations on large-scale topologies, showing that ReWeave outperforms state-of-the-art schemes such as HARP and Flexile in link utilization, response speed, scalability, and packet loss.
\end{itemize}

The remainder of this paper is organized as follows.  
Section~II discusses the existing problems and challenges.  
Section~III presents the motivation through a simplified example.  
Section~IV introduces the framework of ReWeave and elaborates on its components.  
In Section~V, we evaluate the performance of ReWeave.  
A review of related work is presented in Section~VI, followed by the conclusion in Section~VII.

\section{Problem Statement and Challenges}

Path-based TE systems typically follow a two-stage design. Stage one selects forwarding paths based on topology and updates them infrequently due to high overhead. Stage two frequently adjusts traffic splitting ratios on these fixed paths using demand data. This section analyzes the limitations of stage one, explains why they persist in stage two, and outlines challenges in improving path selection.

\subsection{Problem: Static Paths Become Vulnerable Under Failures} \label{sec:problem}
Path selection has a significant impact on the performance and robustness of a TE system.  An ideal set of fixed paths should offer sufficient diversity to ensure high performance,  while keeping path lengths short to minimize latency. Moreover, to support fast failure recovery, the path selection process should be efficient and easy to compute.  
Currently, three commonly used multi-path strategies in TE systems include:

\begin{figure*}[htbp]
    \centering
    \begin{minipage}[t]{0.24\textwidth}
        \centering
        \includegraphics[width=\linewidth]{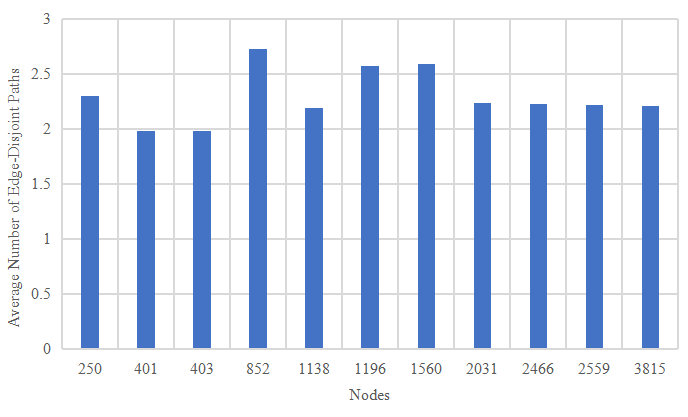}
        \centerline{\small (a) Avg. EDKSP Paths per Pair}
        \label{fig:1a}
    \end{minipage}
    \hfill
    \begin{minipage}[t]{0.24\textwidth}
        \centering
        \includegraphics[width=\linewidth]{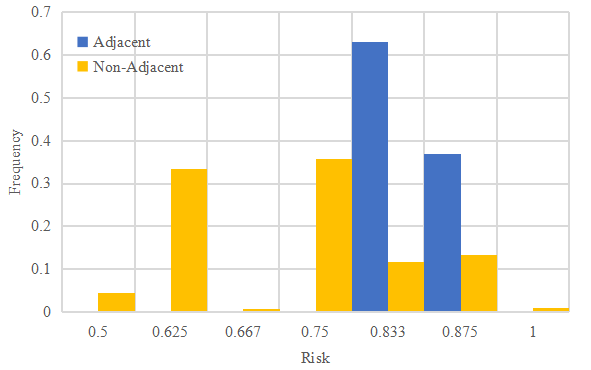}
        \centerline{\small (b) KSP Risk in Viatel}
        \label{fig:1b}
    \end{minipage}
    \hfill
    \begin{minipage}[t]{0.24\textwidth}
        \centering
        \includegraphics[width=\linewidth]{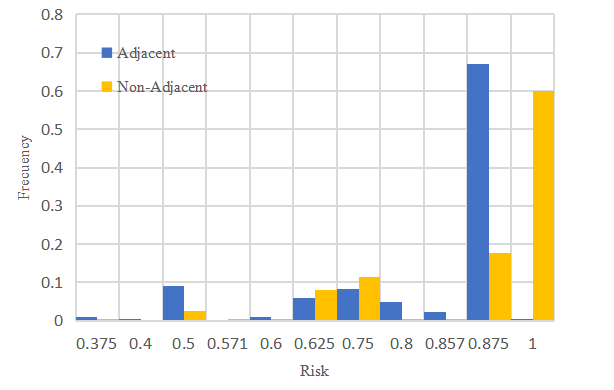}
        \centerline{\small (c) KSP Risk in Cogentco}
        \label{fig:1c}
    \end{minipage}
    \hfill
    \begin{minipage}[t]{0.24\textwidth}
        \centering
        \includegraphics[width=\linewidth]{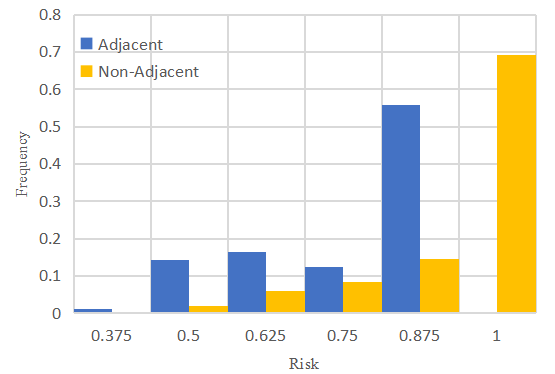}
        \centerline{\small (d) KSP Risk in north\_america}
        \label{fig:1d}
    \end{minipage}
    \caption{(a) Average number of edge-disjoint paths (EDKSP) available per node pair across topologies from TopoHub; (b)-(d) Distribution of maximum edge risk ratio among adjacent vs. non-adjacent node pairs in three backbone topologies (k = 8). Adjacent pairs tend to exhibit lower risk, suggesting higher robustness under failures.}
    \label{fig:risk}
\end{figure*}

\begin{itemize}
    \item \textbf{KSP (k Shortest Paths)}: selects the $k$ shortest paths between a source-destination pair.
    \item \textbf{EDKSP (Edge-Disjoint K Shortest Paths)}: selects $k$ shortest edge-disjoint paths between a source-destination pair.
    \item \textbf{Smore Trees}: constructs randomized routing trees based on Räcke's oblivious routing algorithm\cite{b17}, aiming for path diversity with low latency and low risk.
\end{itemize}

However, none of these strategies fully satisfy our requirements. Smore Trees appear to strike a balance between performance and latency. Unfortunately, as a globally optimized algorithm, its path computation is extremely slow and exhibits scalability bottlenecks as the topology size increases\cite{b28}.

To quantify the robustness of a path set under failures, we adopt the concept of bottleneck risk from \cite{b9}. It is defined as the maximum edge risk ratio within a path set, where an edge's risk ratio is the proportion of paths traversing it. A high bottleneck risk indicates a heavy reliance on a single edge, making the entire path set vulnerable to its failure:


\begin{equation}
\text{Risk}(s,d) = \max_{e \in E} \left( \frac{|\{p \in P_{s,d} : e \in p\}|}{|P_{s,d}|} \right)
\label{eq:risk}
\end{equation}

For EDKSP paths, since all paths are edge-disjoint, each edge appears in exactly one path. Hence, every edge has a risk ratio of $1/|P_{s,d}|$.
Fig.~\ref{fig:risk} shows the average number of edge-disjoint paths (EDKSP) available between node pairs across a range of large-scale backbone networks from TopoHub, where the size of each topology is represented by its number of nodes. We observe that regardless of the value of $k$, most node pairs can only obtain two edge-disjoint paths. This results in a high risk rate of up to 50\%, meaning that if one path fails, only a single alternative remains available. Such limited redundancy makes the network highly vulnerable to failures.

For KSP, we distinguish between adjacent and non-adjacent node pairs. An adjacent node pair has a direct link between them; otherwise, it is non-adjacent. We compute the $k=8$ shortest paths for three topologies used in our evaluation (Section~\ref{sec:V}), and analyze the maximum edge risk ratio across all paths. The results are summarized in Fig.~\ref{fig:risk}. To ensure paths remain short, KSP tends to reuse common edges, turning them into congestion bottlenecks and undermining performance. As the topology size increases, more node pairs exhibit a maximum risk ratio that reaches 1. However, we observe that for adjacent node pairs, the maximum risk ratio remains significantly lower than that of non-adjacent pairs and rarely reaches 1, suggesting that adjacent node pairs are more robust and suitable for load balancing.

Existing TE schemes aim to achieve good performance on a fixed set of paths by allocating different proportions of traffic. However, failure scenarios amplify the drawbacks of each path selection method. This indicates the key bottleneck in fault handling is not the path selection method itself, but rather the static path constraints that limit the solution space, making it impossible to meet performance demands by only adjusting traffic ratios.

\subsection{Challenges} \label{sec:adjustment_latency}

Based on the above analysis, an effective TE solution for failure scenarios must meet the following key requirements:  
\begin{itemize}
  \item[\textit{1)}] Provide a robust set of paths such that load balancing remains feasible even after link failures, preventing significant performance degradation;
  \item[\textit{2)}] Enable fast adjustment upon failure detection to avoid packet loss and service disruption;
  \item[\textit{3)}] Ensure good scalability to handle link failures in large-scale networks.
\end{itemize}

However, designing a TE system that satisfies all these requirements is far from trivial. It faces the following challenges:

\begin{itemize}
  \item[\textit{1)}] On the one hand, we aim to minimize adjustment latency. Localized adjustments achieve this as they do not require the computation of new end-to-end paths or global flow table updates. On the other hand, the remaining paths after failure are often fragile, and without updates, proper load balancing becomes impossible.
    
  \item[\textit{2)}] Fault handling schemes implemented solely at the router level become disconnected from the traffic engineering policies provided by the central controller, leaving the controller unaware of network failures. The algorithms used to calculate traffic distribution proportions, which are the core design of most TE systems, often rely on machine learning methods. These TE models are trained on a fixed set of paths. When paths change due to failures, the performance of the original schemes degrades, and retraining these models is time-consuming\cite{b15}.
\end{itemize}

\section{Our Insight} \label{sec:III}
\subsection{Theoretical Rationale for Localized Rerouting}

To formalize the intuition that localized rerouting can outperform global rerouting, we provide a theoretical analysis by modeling the traffic increase ($\Delta f$) on an arbitrary link $e_0$ after a failure. Let $F_{sd}^{\text{fail}}$ be the amount of traffic on a path between source $s$ and destination $d$ that is disrupted by a failure.

For baseline schemes that rely on source-level rerouting, this traffic is redistributed among the surviving end-to-end paths. The expected traffic increase on link $e_0$ is the sum of all redistributed traffic, weighted by the proportion of surviving paths traversing $e_0$. Consistent with the edge risk ratio defined in Sec. II-A, we term this proportion the Residual Risk ($\text{Risk}_{\text{remain}}$).
This leads to:

\begin{equation}
\Delta f_{e_0}^{\text{base}} = \sum_{(s,d)} \left( F_{sd}^{\text{fail}} \cdot \text{Risk}_{\text{remain}}^{(s,d)}(e_0) \right)
\end{equation}

In contrast, ReWeave reroutes the total failed traffic volume locally using a pre-computed set of backup paths. The traffic increase on link $e_0$ is therefore the total failed volume multiplied by the proportion of backup paths that traverse $e_0$, which we term the Backup Risk ($\text{Risk}_{\text{backup}}$):

\begin{equation}
\Delta f_{e_0}^{\text{reweave}} = \text{Risk}_{\text{backup}}(e_0) \cdot \left( \sum_{(s,d)} F_{sd}^{\text{fail}} \right)
\end{equation}

The superiority of our approach stems from the fact that $\text{Risk}_{\text{backup}}$ is inherently lower and more distributed. As shown in our analysis in Fig.~\ref{fig:risk}, the path sets between adjacent nodes (our backup paths) exhibit significantly lower maximum edge risk compared to the end-to-end KSP paths used by baselines, whose $\text{Risk}_{\text{remain}}$ is often concentrated on a few bottleneck links. This theoretical difference explains why ReWeave achieves lower MLU in practice.

\begin{figure*}[htbp]
    \centering
    \begin{minipage}[t]{0.24\textwidth}
        \centering
        \includegraphics[width=\linewidth]{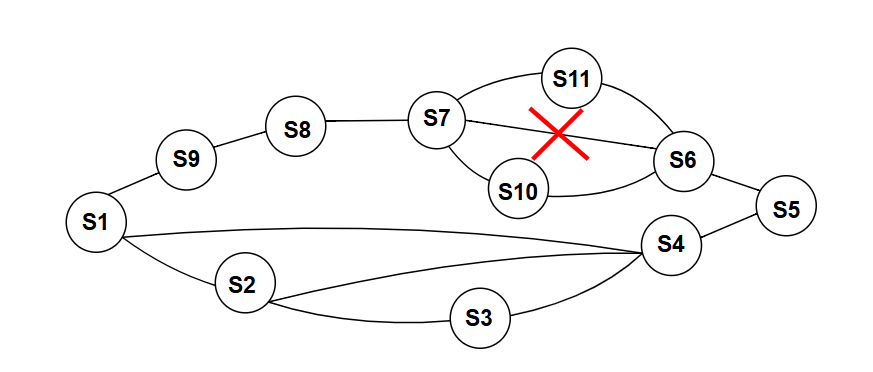}
        \centerline{\small (a) Illustrative Topology}
        \label{fig:3a}
    \end{minipage}
    \hfill
    \begin{minipage}[t]{0.24\textwidth}
        \centering
        \includegraphics[width=\linewidth]{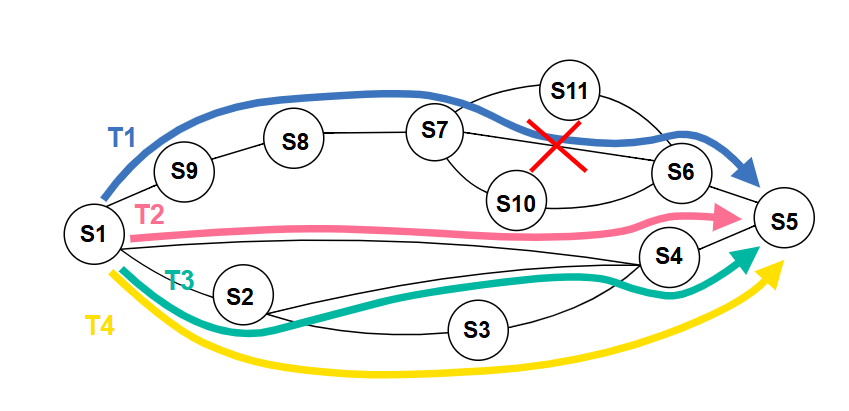}
        \centerline{\small (b) KSP Routing}
        \label{fig:3b}
    \end{minipage}
    \hfill
    \begin{minipage}[t]{0.24\textwidth}
        \centering
        \includegraphics[width=\linewidth]{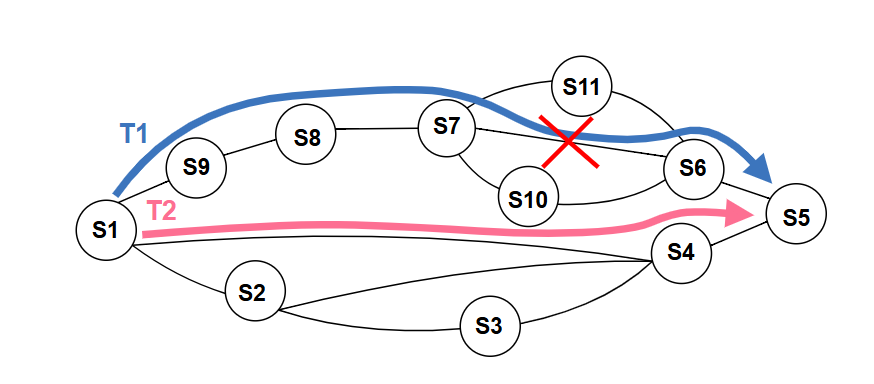}
        \centerline{\small (c) EDKSP Routing}
        \label{fig:3c}
    \end{minipage}
    \hfill
    \begin{minipage}[t]{0.24\textwidth}
        \centering
        \includegraphics[width=\linewidth]{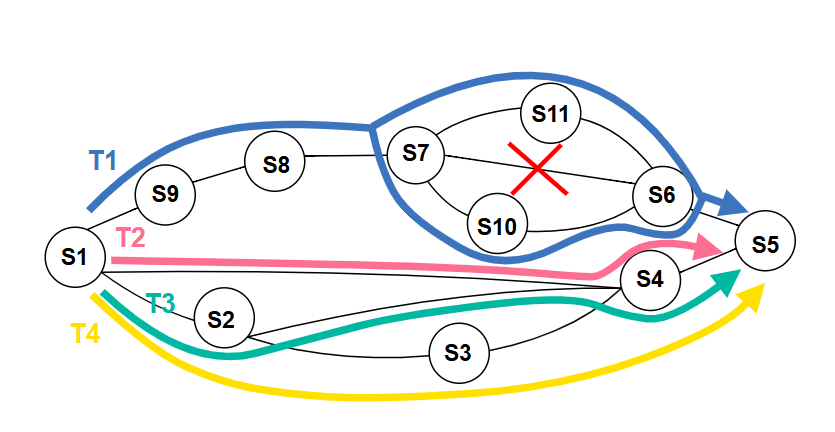}
        \centerline{\small (d) ReWeave}
        \label{fig:3d}
    \end{minipage}
    \caption{Example topology and path behaviors under single-link failure: (a) illustrative topology; (b) behavior of KSP routing; (c) behavior of EDKSP routing; (d) behavior of our proposed scheme.}
    \label{fig:topo_all}
\end{figure*}

\subsection{Illustrative Topology}

To provide a concrete illustration of the theoretical principles discussed above, we now consider how to address the challenges associated with implementing local dynamic paths. We start with a simple example under a single-link failure scenario, which is a common setting to evaluate the robustness of TE schemes.

Fig.~\ref{fig:topo_all}(a) shows our example topology, which consists of 11 switches and 17 links. All links have a unit capacity. The traffic demand from $S_1$ to $S_5$ is 1.2. Suppose link $(S_7, S_6)$ fails. We use $k=4$ as the number of paths per source-destination pair. For simplicity, we assume uniform splitting among tunnels.

Fig.~\ref{fig:topo_all}(b) shows the case of KSP routing. In the normal state, four tunnels $T_1$, $T_2$, $T_3$, and $T_4$ are available, each carrying 25\% of the traffic. When the failure occurs, tunnel $T_1$ becomes unavailable. The 25\% traffic on $T_1$ is redistributed to the remaining tunnels. However, regardless of how the redistribution is done, all traffic now passes through bottleneck link $(s_4, s_5)$, which becomes congested as the total flow exceeds its capacity ($1.2 > 1$).

Fig.~\ref{fig:topo_all}(c) shows the case of EDKSP routing. In the normal state, at most two tunnels $T_1$ and $T_2$ are provided, each carrying 50\% of the traffic. When $T_1$ fails, all traffic is forced onto $T_2$, causing congestion on all its links.

\begin{figure}[htbp]
  \centerline{\includegraphics[width=0.48\textwidth]{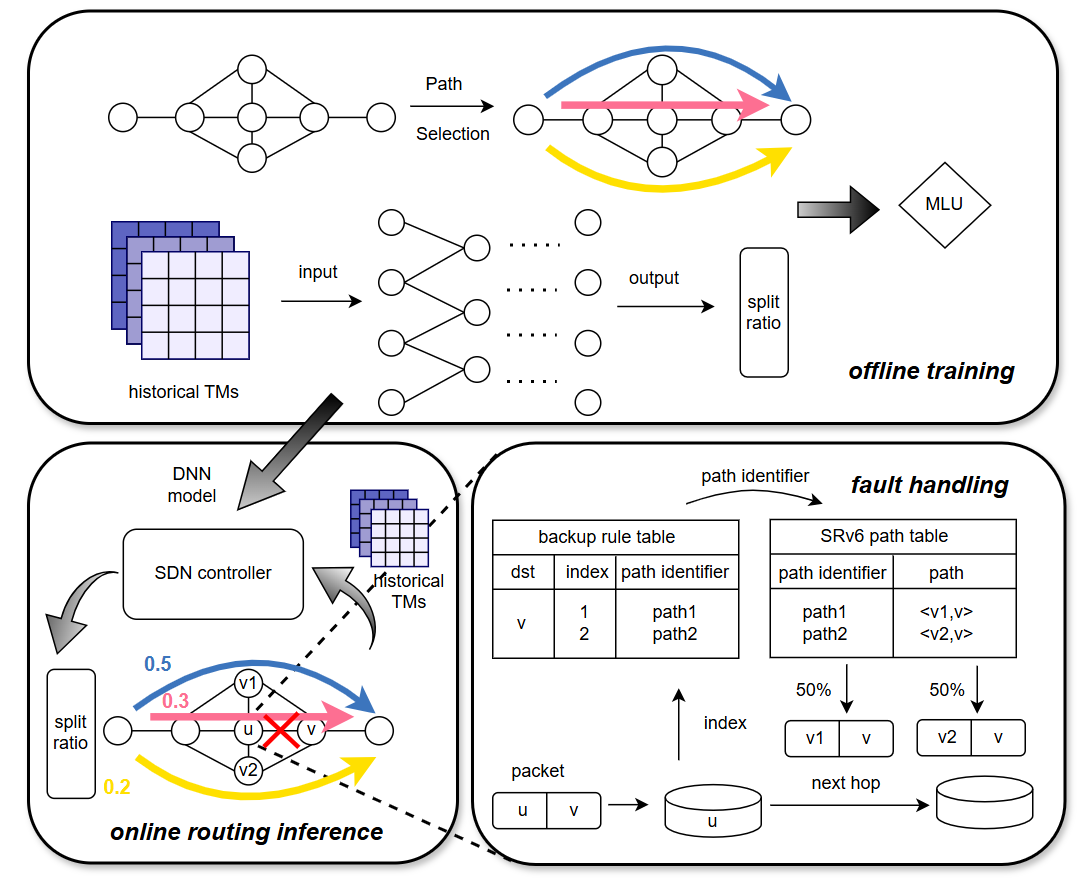}}
  \caption{Architecture of ReWeave}
  \label{fig:architecture}
\end{figure}
\subsection{Design Intuition and Validation}
Our design principle is to maintain logical path stability by minimizing modifications, which is the key to ensuring our ML-based optimizer can continue to effectively minimize MLU post-failure. When links fail, TE models falter if the path structure changes too drastically, invalidating their pre-trained policies. ReWeave's local weaving, illustrated in Fig.~\ref{fig:topo_all}(d), directly addresses this challenge. 
Upon failure of link ($S_7,S_6$), the upstream router $S_7$ splits the traffic that was originally forwarded through the failed link into two alternative sub-paths. Logically, the tunnel $T_1$ from $S_1$ to $S_5$ (originally \{$S_1,S_9,S_8,S_7,S_6,S_5$\}) is now replaced with two sub-tunnels: \{$S_1,S_9,S_8,S_7,S_{11},S_6,S_5$\} and \{$S_1,S_9,S_8,S_7,S_{10},S_6,S_5$\}.
From the perspective of source $S_1$, tunnel $T_1$ remains logically available. This is the crucial insight: by preserving the tunnel's logical integrity, the pre-trained ML model's policy---in this case, allocating 25\% of traffic---remains valid and effective without causing congestion. Therefore, minimizing path modification is not an end in itself but the fundamental enabler for continuously pursuing the primary objective of minimizing MLU.


\section{Framework and Design}

\subsection{TE Model} \label{sec:IV-A}

\renewcommand{\arraystretch}{1.3}
\begin{table}[h]
  \centering
  \caption{Definition of notations}
  \begin{tabular}{lp{6.5cm}}
  $G=(V, E)$ & The network topology represented as a graph $G$, where $V$ is the node set and $E$ is the edge set. \\
  $c(e)$ & The capacity of link $e$. \\
  $D_t^{|V|\times|V|}$ & The demand matrix at time $t$, where the element $d_{ij}$ in the $i$-th row and $j$-th column represents the traffic demand from node $i$ to node $j$ at time $t$. \\
  $\mathcal{R}_t$ & The overall TE configuration, represented by the traffic allocation ratio matrix at time epoch $t$. \\
  $P_{s,d}$ & The set of all paths from source $s$ to destination $d$. \\
  $\lambda_p$ & The traffic split ratio on path $p$. \\
  $f_e$ & The traffic passing through link $e$.
  \end{tabular}
  \label{tab:notations}
  \end{table}
We introduced the mathematical notations and definitions of the TE model in Table \ref{tab:notations}. 
Given the demand matrix $D_t$ and the ratio configuration $\mathcal{R}_t$ at time $t$, the traffic allocated on each link can be calculated. The objective of TE is to minimize the maximum link utilization, subject to the constraints of capacity and allocation ratio.
\begin{align}
  \underset{\mathcal{R}_t}{\text{minimize}} \quad & MLU(\mathcal{R}_t, D_t) \label{eq:MLU-1} \\[-1mm]
  \text{subject to} \quad & \sum_{p \in P_{s,d}} \lambda_p = 1, \quad \forall s,d \in V, s \neq d \\[-1mm]
  & \frac{f_e}{c(e)} \leq MLU, \quad \forall e \in E \\[-1mm]
  & f_e = \sum_{\substack{s,d \in V,\, p \in P_{s,d},\, e \in p}} D_{sd} \cdot \lambda_p, \quad \forall e \in E
\end{align}


\subsection{Path Selection Methods} \label{sec:IV-B}

While alternative strategies like SMORE [17] offer theoretical robustness, we adopt a shortest-path-based approach for two practical reasons: its superior computational efficiency and its direct alignment with our local repair mechanism, which leverages the structural properties of shortest paths.
For each node pair in the network, we first compute its $k$ shortest edge-disjoint paths. The specific method involves recursively calculating the shortest path in the current network and then removing all edges belonging to that path from the topology. This process is repeated until the source and destination nodes become disconnected or the number of generated paths reaches $k$. For each pair of adjacent nodes, we compute the shortest $k$ paths between them as backup paths.

This routing strategy offers two key advantages. First, for each pair of adjacent nodes, the computed path set tends to avoid shared links, thereby enhancing fault tolerance. As shown in the Table~\ref{tab:path-selection}, with the increase in network scale, nearly every link between adjacent node pairs is associated with $k$ backup paths. This ensures multiple alternative routing options for traffic upon a link failure, thereby enhancing the robustness of the overall path set. Second, the Table~\ref{tab:path-selection} also reveals that the number of edge-disjoint paths is often less than $k$, naturally limiting the backup paths below $k$ and resulting in lower overhead compared to schemes installing $k$ shortest paths for all node pairs. This design effectively reduces the cost of maintaining backup paths. We will compare our path selection and backup strategy with other routing methods in Section V to demonstrate its suitability and superiority.
\begin{table}[ht]
    \centering
    \caption{Comparison of EDKSP and KSP Paths \& Coverage}
    \begin{tabular}{|c|c|c|c|c|}
        \hline
        \textbf{Topology} & \textbf{EDKSP} & \textbf{backup} & \textbf{KSP} & \textbf{P(8-backup)} \\
        \hline
        Viatel & 15095 & 1240 & 58892 & 36.96\% \\
        Cogentco  & 52869 & 3194 & 307590 & 87.26\% \\ 
        north\_america   & 133042 & 5440 & 458880 & 100.00\% \\ 
        africa  & 322990 & 8336 & \textgreater 400000
 & 100.00\% \\ 
        \hline
    \end{tabular}
    \label{tab:path-selection}
\end{table}

\subsection{TE Algorithm for Traffic Ratio Assignment} \label{sec:IV-C}

The core problem of TE is to determine the traffic ratio on each path such that the resulting MLU is minimized, given a path set and a traffic demand matrix observed during a fixed measurement interval (see \eqref{eq:MLU-1}).In online deployment, future demand matrices are not known in advance. One common approach is to predict future traffic demand based on historical measurements\cite{b29} and then solve the TE model as if the predicted demand were real. However, this two-stage approach introduces compounded prediction error due to the coupling between prediction and optimization\cite{b28}. Instead, inspired by \cite{b28}, we adopt a direct approach to predict the future optimal traffic ratio. We design a five-layer fully connected neural network that learns the mapping from past demand matrices to future optimal path ratios. 
The new TE objective is then:

\begin{equation}
  \underset{\mathcal{R}_t(D_1, \dots, D_{t-1})}{\text{minimize}} \quad MLU(\mathcal{R}_t(D_1, \dots, D_{t-1}), D_t)\label{eq:MLU-2}
\end{equation}

The output of the DNN is then post-processed via matrix operations to obtain path ratios compliant with the TE model.

The reason for selecting a fully connected network is twofold: 
First, many existing ML-based TE schemes rely on assumptions of fixed topology, and are thus limited to lightweight local rerouting mechanisms\cite{b28,b29}. Our approach relaxes this assumption while achieving more robust global optimization. 
Second, compared to topology-specific GNNs or custom architectures\cite{b3,b15}, the use of a plain, interpretable DNN highlights the universality and compatibility of our proposed scheme.

\subsection{Fault Handling Scheme}

Our fault handling scheme operates entirely in the data plane, independent of the central controller. It utilizes the traffic allocation ratios computed by the ML model under fault-free conditions, which are held constant upon a failure. This ensures recovery is immediate and localized, without altering traffic from unaffected flows. For our evaluation, we simulate this traffic adjustment based on the current topology and link failures. The following subsections detail the practical deployment of this scheme.


\subsubsection{Router Structure}

We employ SRv6 tunnels to implement proportional traffic distribution and failure-aware traffic adjustment across multiple paths. Routers maintain rule tables for splitting traffic and SRv6 path tables for storing end-to-end paths\cite{b21}. 
In an SDN network, where traffic demands are considered between every pair of distinct routers, each router in a network of $N$ routers needs to install $M(N-1)$ forwarding rules to support hash-based traffic splitting towards the other $(N-1)$ routers. Furthermore, assuming each router has a degree $d$, an additional $Md$ backup entries must be reserved to handle potential failures on the $d$ adjacent links.
Here, $M$ is the granularity of traffic splitting; the larger the $M$, the finer and more accurate the splitting. This reduces the deviation between actual and calculated traffic ratios and ensures traffic engineering results closely match the algorithm's output. Each rule entry consists of 8 bytes: the first 4 bytes store matching fields, and the remaining 4 bytes store an action field that holds a path identifier. The SRv6 path table maps each path identifier to an end-to-end path comprising $L$ SIDs, where $L$ depends on the maximum path length.

\subsubsection{Link-Level Failure Detection}

Conventional path-level failure detection establishes monitoring sessions\cite{b22} between each source-destination pair. These sessions piggyback probe packets on existing traffic, unlike the "Hello" exchange mechanism used for link-level failure detection. This implicit probing introduces no extra overhead and can detect more types of failures\cite{b23}. In SDN networks, each link is naturally considered as a separate path between its two endpoint routers. Thus, a monitoring session is created for each link, allowing both endpoint routers to detect link failures.

\subsubsection{State-Independent Adjustment Strategy}

Consider a faulty link $e = (u,v)$. Since $(u,v)$ forms a node pair with traffic demand, the TE system preconfigures multiple paths from $u$ to $v$ during the preparation phase. 

When the affected flow reaches the edge router, the packet that was originally supposed to be forwarded to the next-hop node \( v \) can no longer proceed along the failed link. The router \( u \) identifies node \( v \) as the other endpoint of the failed link and randomly selects a path from its locally stored backup routing table for destination \( v \), with equal probability. The corresponding Segment ID (SID) list of the selected path is then inserted into the packet header, enabling detour forwarding around the failure.
This mechanism fully demonstrates the flexibility of Segment Routing over IPv6 (SRv6): SDN routers can dynamically modify the SID list of a packet, enabling fault recovery through local control without the need to recompute the entire path.
In the rare event that no local backup paths are available between the endpoints of a failed link, the failure is handled by established upstream recovery mechanisms, such as source-level rerouting, ensuring comprehensive fault tolerance.
With this design, we achieve fast rerouting after link failures, ensuring service continuity and enhancing network robustness.










\section{Evaluation} \label{sec:V}
\subsection{Preliminary}

\subsubsection{Path Selection Algorithm}


We aim to compute $k=8$ edge-disjoint shortest paths between any two nodes. 
Regarding the choice of $k=8$, we performed a sensitivity analysis by evaluating the average $\text{Risk}_{\text{backup}}$ for $k$ ranging from 6 to 10. The results, as shown in Table\ref{tab:k_sensitivity}, indicate that the risk metric is relatively stable within this range across all topologies. As simply increasing $k$ does not guarantee a monotonic risk reduction and incurs higher state overhead, $k=8$, also used by our baseline DOTE\cite{b28}, represents a reasonable and well-justified choice for our experiments.
Furthermore, to validate the rationale of our proposed routing method and backup strategy, we compare its performance with an alternative routing method applied to ReWeave: KSP, which generates $k$ shortest paths using Yen's algorithm.
\begin{table}[h!]
    \centering
    \caption{Sensitivity Analysis of Average $\text{Risk}_{\text{backup}}$ for Different $k$}
    \label{tab:k_sensitivity}
    \begin{tabular}{|l|c|c|c|c|c|}
        \hline
        \textbf{Topology} & \textbf{k=6} & \textbf{k=7} & \textbf{k=8} & \textbf{k=9} & \textbf{k=10} \\
        \hline
        Viatel        & 0.4843       & 0.4936       & 0.4698       & 0.4800       & 0.4841        \\
        Cogentco      & 0.4274       & 0.3971       & 0.3791       & 0.3612       & 0.3367        \\
        north\_america & 0.4078       & 0.3798       & 0.3577       & 0.3402       & 0.3233        \\
        africa        & 0.4356       & 0.4095       & 0.3907       & 0.3683       & 0.3495        \\
        \hline
    \end{tabular}
\end{table}

\subsubsection{DNN Architecture}

Our deep neural network (DNN) architecture, drawing from the empirically validated optimal design in DOTE\cite{b28}, comprises five fully connected layers, each hidden layer featuring 128 neurons. For activation functions, all hidden layers employ ReLU(x), while the output layer utilizes Sigmoid(x). 


\subsubsection{Topology}

In our evaluation, we employ two real-world network topologies, Viatel and Cogentco,  from the Topology Zoo\cite{b24}, along with two synthetic topologies, north\_america and africa, obtained from TopoHub\cite{b25}. These topologies are arranged in increasing scale, representing the anticipated trend in the construction of future backbone networks. Table \ref{tab:topologies} presents the detailed information for these topologies, including the number of nodes, and the number of links. We recursively remove all one-degree nodes from the topology to ensure that the network does not become disconnected due to a single link failure.

\begin{table}[b]
  \caption{Network Topologies Used in Evaluation}
  \begin{center}
  \begin{tabular}{|c|c|c|c|c|}
  \hline
  \textbf{Topology}&\textbf{Nodes}&\textbf{Links} \\
  \hline
   Viatel & 88 & 184  \\
  \hline
   Cogentco & 197 & 486  \\
  \hline
  north\_america & 250 & 700  \\
  \hline
  africa & 403 & 1072  \\
  \hline
  \end{tabular}
  \label{tab:topologies}
  \end{center}
  \end{table}

\subsubsection{Traffic Demand}

For the network topologies selected from TopoHub, we generate 200 synthetic traffic matrices using the gravity model\cite{b19} to simulate realistic traffic demand distributions in practical network environments. For each WAN, we take the first 75\% of the TMs as the training set, and the remaining 25\% as the testing set.

\subsubsection{Baseline}

We compare \textbf{ReWeave} with the following baselines to demonstrate its advantages:

\begin{itemize}
    \item \textbf{Linear Programming}: The theoretical optimal solution 
    obtained with knowledge of future demands and failures, 
    serving as a normalized reference for evaluation.

    \item \textbf{DOTE}\cite{b28}: A deep learning-based TE method 
    that predicts TE configurations directly from historical traffic demands. 
    Failures are handled via source node rerouting.

    \item \textbf{FIGRET}\cite{b29}: Another deep learning-based TE method 
    that introduces fine-grained robustness constraints 
    to avoid excessive link loads. 
    This mitigates performance degradation under failures, 
    also using source node rerouting.

    \item \textbf{HARP}\cite{b15}: A TE method designed for topology changes, 
    which generates TE configurations from predicted traffic matrices. 
    Its RAU module moves flows away from low-capacity tunnels 
    to maintain performance. 
    Upon failures, it relearns the topology and produces new allocation ratios.

    \item \textbf{Flexile}\cite{b14}: A linear-programming-based TE approach. 
    In the offline phase, it pre-identifies critical failure scenarios per flow. 
    During failures, critical flows are prioritized, 
    while bandwidth is allocated to non-critical flows 
    to meet demands as much as possible.
\end{itemize}

\subsubsection{Evaluation Metrics}

Maximum Link Utilization (MLU) is a classical metric for evaluating the load-balancing performance of TE schemes\cite{b15}. We adopt the normalized MLU as our primary evaluation metric, which reflects the gap between our method and the omniscient optimal solution obtained via linear programming (denoted as $\text{MLU}_{\text{LP}}$). The closer the normalized MLU is to 1, the better the TE performance. The computation is as follows:
$(\text{normalized})\ \text{MLU} = \frac{\text{MLU}}{\text{MLU}_{\text{LP}}}$


Additionally, we adopt the same optimization target as Flexile—the 99th percentile of bandwidth loss (PercLoss)—to evaluate the worst-case performance of TE schemes. The specific calculation is as follows:
\begin{align}
  \text{loss} &= 1 - \frac{1}{\min(1, \text{MLU})} \\[-1mm] 
  \text{PercLoss} &= \max_{s,d \in V} \text{FlowLoss}(f_{sd}, \beta)
\end{align}
Here, given the computed MLU, the actual flow from source $s$ to destination $d$ is scaled to $d_{sd} / \min(1, \text{MLU})$, yielding the corresponding loss rate. We define $\text{FlowLoss}(f_{sd}, \beta)$ as the loss rate of flow $f_{sd}$ in the scenarios whose cumulative probability exceeds $\beta$. Then, $\text{PercLoss}$ is the maximum of all such flow losses, indicating worst-case bandwidth degradation.
  
  


\subsection{Single-Link Failure Scenarios}

\begin{figure}[htbp]
\centerline{\includegraphics[width=\columnwidth]{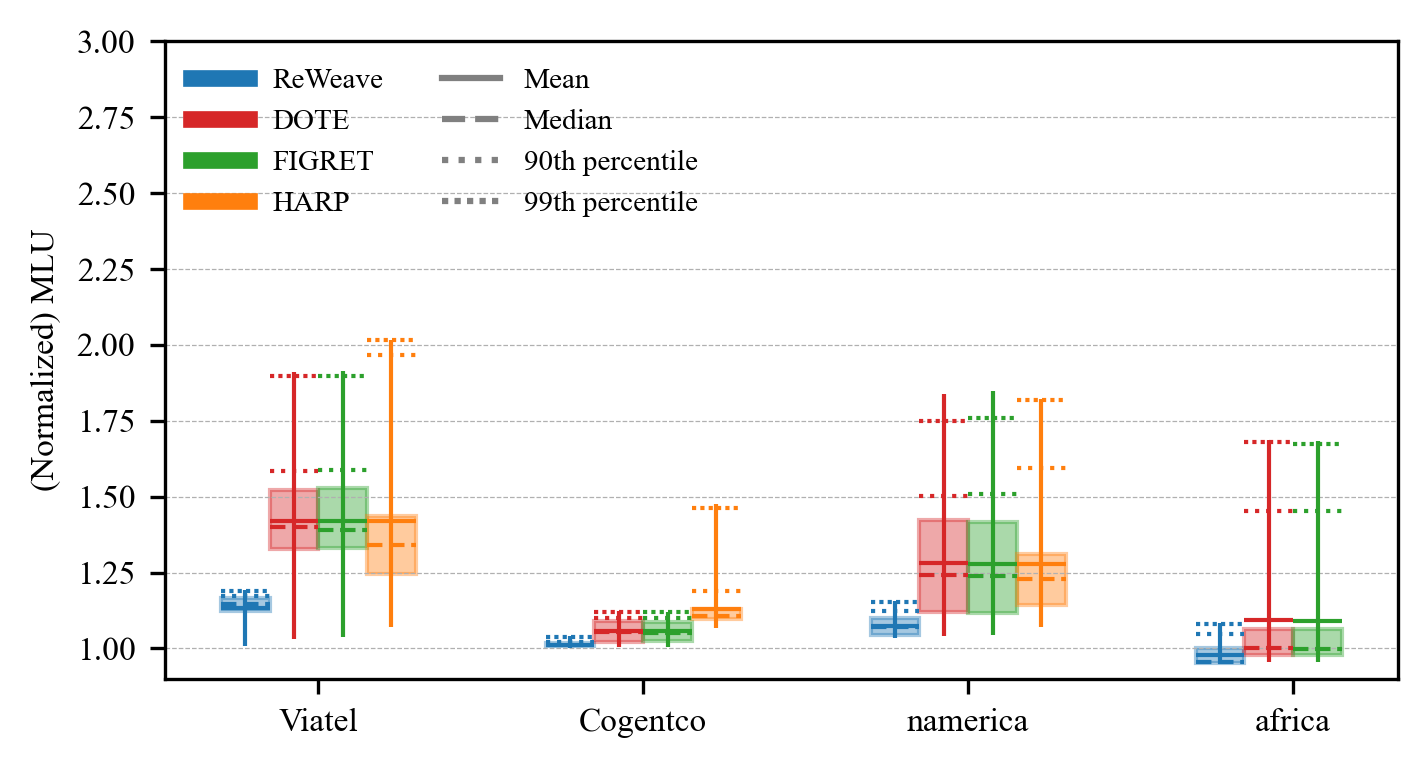}}
\caption{Normalized MLU under single-link failures  
across four topologies.}
\label{fig:mlu_single_failure}
\end{figure}

Single-link failures are the most frequently observed failures in real-world networks. To simulate this scenario, for each test traffic matrix (TM), we randomly select a link index and set its capacity to zero, effectively emulating a link failure. As shown in Fig.~\ref{fig:mlu_single_failure}, we compare ReWeave with baseline methods on four medium-sized network topologies in terms of normalized maximum link utilization (MLU). We observe that ReWeave consistently achieves the best performance. Specifically:

\textbf{DOTE}: Although it integrates a traditional router-level rerouting mechanism, it only reacts to failure-affected node pairs. These flows compete with unaffected flows for the remaining resources, worsens congestion. In contrast, ReWeave significantly alleviates such pressure and lowers MLU.

\textbf{FIGRET}: This method applies fine-grained constraints at the controller level to balance bursty traffic and reduce fluctuations. However, the controller is unaware of link failures, and thus lacks robustness when faults occur.

\textbf{HARP}: By leveraging a GNN structure, HARP detects topological changes and achieves relatively low average MLU. However, fragile residual paths and larger network sizes hinder convergence to optimal solutions. HARP often produces suboptimal results and consumes more resources. Given a 1-day time and 96GB memory limit, it fails on the africa topology, revealing scalability limitations.

\textbf{ReWeave}: By employing simple backup paths, our method addresses the fragility of residual paths after failures. Compared to DOTE and FIGRET, ReWeave reduces the average MLU by 4.1\%--20.2\%, and the worst-case MLU by 7.2\%--37.8\%. Even compared with HARP, which has global fault-awareness, ReWeave achieves comparable or superior performance, reducing average MLU by 10.5\%--20.1\% and worst-case MLU by 29.5\%--40.9\%. Notably, on the africa topology, ReWeave even outperforms the optimal solution obtained via linear programming. This suggests that fixed paths restrict the solution space, while our local adjustments enable escaping from local optima.

\begin{figure}[htbp]
    \centering
    \begin{minipage}[t]{0.48\columnwidth}
        \centering
        \includegraphics[width=\linewidth]{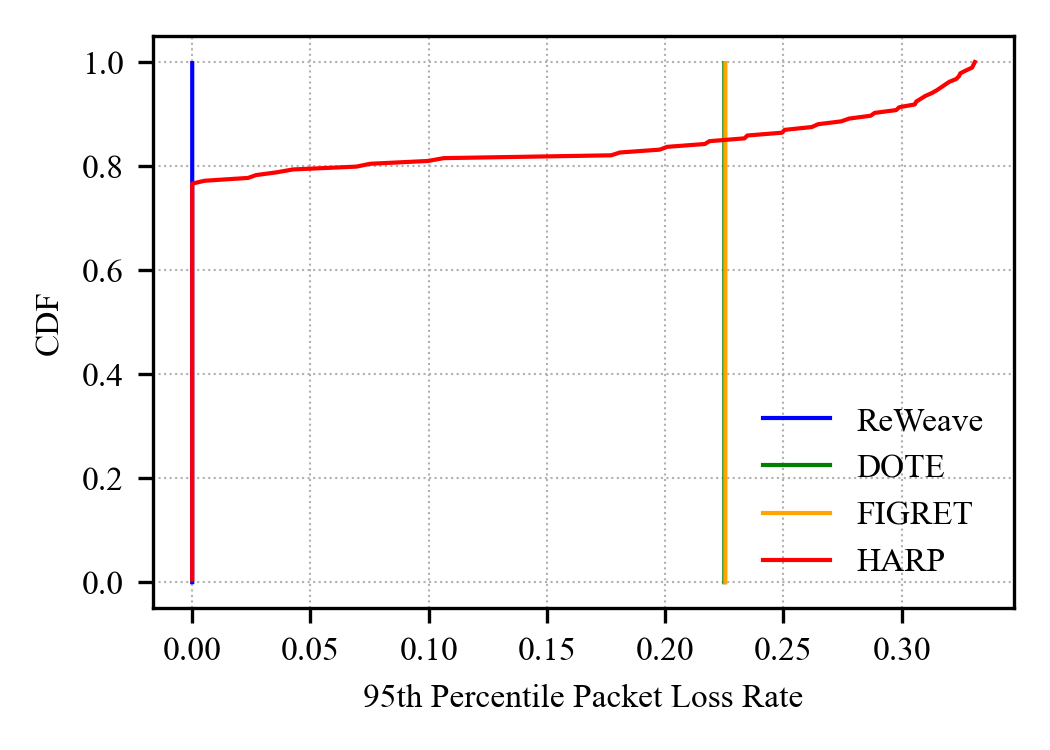}
        \centerline{\small (a) Viatel}
    \end{minipage}
    \hfill
    \begin{minipage}[t]{0.48\columnwidth}
        \centering
        \includegraphics[width=\linewidth]{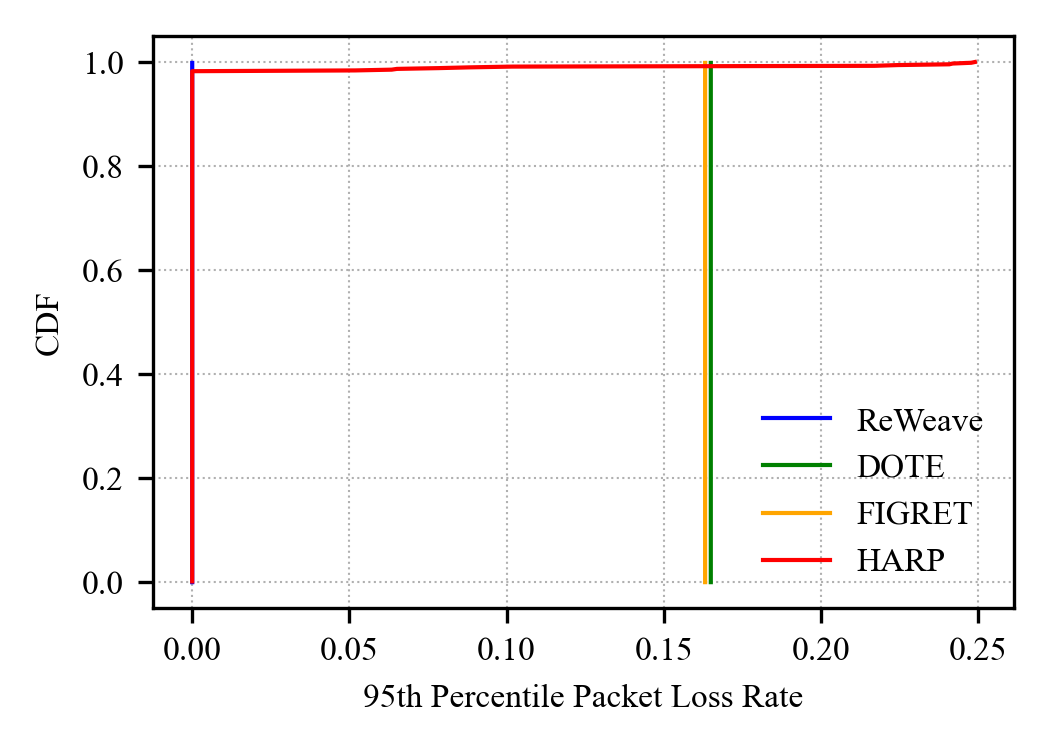}
        \centerline{\small (b) north\_america}
    \end{minipage}
    \caption{90th percentile packet loss rate comparison. Flexile achieves zero packet loss across all topologies and is used as a baseline.}
    \label{fig:loss_90_percentile}
\end{figure}

We also compare the packet loss rates of all schemes, using the Flexile-optimized result (0\% packet loss) as the baseline. Since all schemes exhibit zero 90th percentile packet loss on the Cogentco and Africa topologies, we focus on the Viatel and north\_america topologies. As shown in Fig.~\ref{fig:loss_90_percentile}, we present the 90th percentile packet loss rates of each scheme on both topologies, with Flexile serving as the reference.

DOTE and FIGRET sacrifice congestion control to achieve rapid failure recovery, leading to situations where link utilization exceeds 100\%, resulting in packet loss. Although HARP uses controller-based adjustments to reallocate traffic away from affected paths, it still suffers from congestion in some cases, where a large amount of traffic is redirected to a single link. This behavior stems from the fragility of the remaining paths after link failures. In contrast, our approach effectively augments the set of available paths using backup routes, distributing traffic more evenly, keeping utilization under control, and ultimately reducing packet loss.

\subsection{Multi-Link Failure Scenarios}

\begin{figure}[htbp]
    \centering
    \begin{minipage}[t]{0.48\columnwidth}
        \centering
        \includegraphics[width=\linewidth]{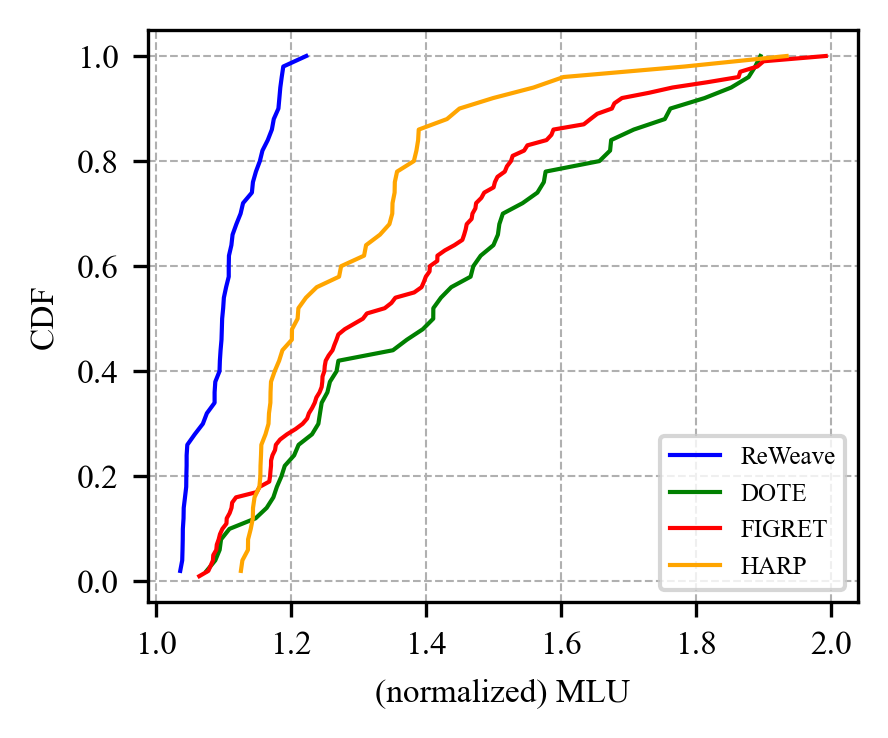}
        \centerline{\small (a) Two Link Failures}
    \end{minipage}
    \hfill
    \begin{minipage}[t]{0.48\columnwidth}
        \centering
        \includegraphics[width=\linewidth]{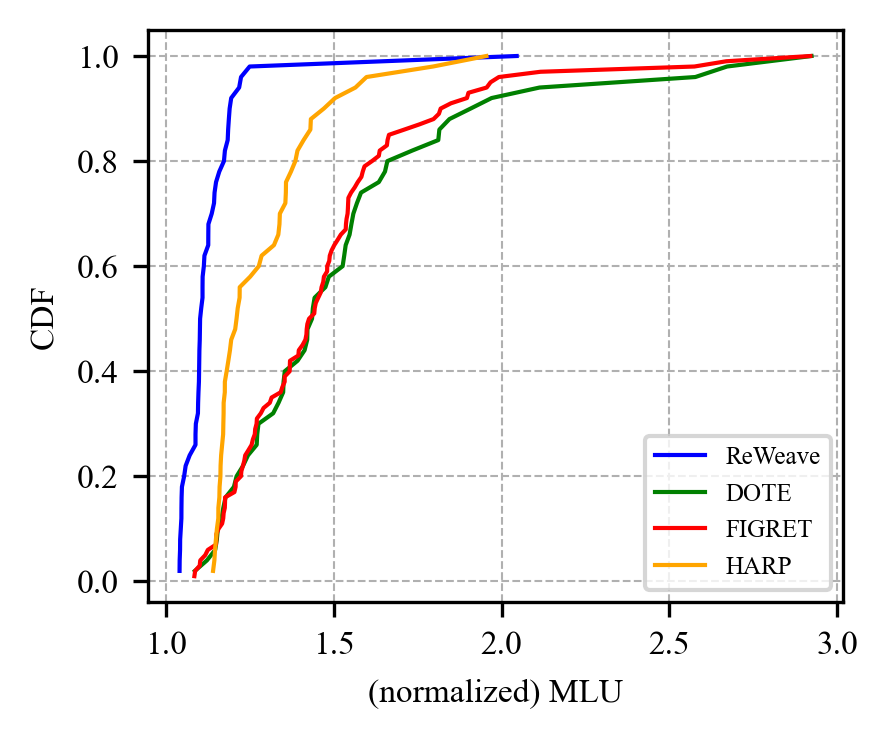}
        \centerline{\small (b) Three Link Failures}
    \end{minipage}
    \caption{CDF of normalized link utilization under multi-link failure scenarios on the north\_america topology. }
    \label{fig:multi_link_failures}
\end{figure}

To evaluate the robustness of our approach under multiple link failures, we conduct experiments on the north\_america topology by injecting two and three simultaneous link failures. For each experimental run, the set of failed links was chosen uniformly at random from all links in the topology. The results are shown in Fig.~\ref{fig:multi_link_failures}. Our approach consistently delivers low-congestion solutions in the presence of multiple failures and significantly outperforms all baselines across scenarios. In the case of two link failures, our scheme maintains normalized link utilization below 1.3 in all settings, ensuring stable traffic engineering performance even under failures. Even with three concurrent link failures, our method achieves normalized utilization below 1.5 in 99\% of the cases, demonstrating substantially greater robustness compared to other approaches.

\subsection{Overhead Analysis}
The primary overheads of our scheme are the offline precomputation of paths and the online state storage, which we analyze in this section. The data plane rerouting time itself, which involves modifying a packet's SRv6 SID list, is negligible. Regarding state, since ReWeave provisions backup paths on a per-link basis rather than all node-pairs, the space overhead scales efficiently with the number of links ($O(e)$).


\subsubsection{Precomputation Time}

Table~\ref{tab:precomputation_time} presents the precomputation time of each TE scheme. For machine learning-based methods, this refers to the time required for model training. In the case of the protection-based scheme Flexile, it represents the offline optimization time based on available traffic matrices and failure probabilities. 
Benefiting from a simple neural network architecture, our method achieves a training time comparable to that of DOTE, and significantly faster than FIGRET, which involves additional sensitivity constraints. Due to the complexity of its GNN-based architecture, HARP requires 2--3$\times$ the training time of our method. Flexile exhibits rapidly increasing offline computation time as network size grows, and neither Flexile nor HARP can complete precomputation within one day for the africa topology.
Notably, our method does not rely on specific failure scenarios or topological features, making its precomputation time particularly advantageous for large-scale networks. This demonstrates that our scheme has low retraining cost and excellent scalability, enabling quick adaptation to evolving topologies.

\subsubsection{Computation Time}

Table~\ref{tab:precomputation_time} also compares the computation time of each TE scheme in response to failures, specifically the time taken to recompute traffic allocations for the next time interval. The term pred-TM denotes the time required to predict the traffic demand of the upcoming interval based on historical data. For prediction-based approaches such as LP and HARP, the total computation time includes both the prediction and the execution of the TE algorithm.
In contrast, ReWeave and DOTE perform rerouting directly at the failure-edge or source routers, respectively, without relying on a centralized controller. As a result, their response times are extremely fast and essentially equivalent to model inference latency. FIGRET exhibits slower inference due to sensitivity constraints, as already reflected in its precomputation stage, though its performance remains within the same order of magnitude.
HARP suffers from a significantly higher computational cost due to its complex neural architecture. After a failure, the central controller requires more time to reallocate traffic from the affected links, severely impacting scalability in large topologies.
The traditional LP approach offers consistent runtime but is clearly outperformed by learning-based methods in terms of speed. These results emphasize that rapid failure recovery in large-scale networks should be executed at the router level, avoiding the delays introduced by centralized failure detection and recomputation.

\begin{table*}[ht]
  \caption{Precomputation and Failure Response Time (in seconds) across Different Topologies}
  \begin{center}
  \begin{tabular}{|c|c|c|c|c|c|c|c|c|c|c|}
    \hline
    \textbf{Topology} 
    & \multicolumn{5}{c|}{\textbf{Precomp. Time (s)}} 
    & \multicolumn{5}{c|}{\textbf{Comp. Time (s)}} \\
    \hline
    & Flexile & HARP & FIGRET & DOTE & ReWeave 
    & LP & HARP & FIGRET & DOTE & ReWeave \\
    \hline
    Viatel & 412 & 5507& 3516 & 1195 & 1094 & 6.54 & 5.34 & 0.07 & 0.02 & 0.02  \\
    Cogentco & 1316 & 14435 & 7455 & 2893 & 2634 & 19.82 & 15.22 & 0.28 & 0.06 & 0.06 \\
    north\_america & 16958 & 22279 & 5679 & 3156 & 3050 & 55.86 & 35.81 & 1.00 & 0.15 & 0.15  \\
    africa &  - & - & 19275 & 8020 & 7707 & 142.66 & - & 1.67 & 0.28 & 0.29  \\
    \hline
  \end{tabular}
  \label{tab:precomputation_time}
  \end{center}
\end{table*}

\subsubsection{Average End-to-End Delay under Link Failure}

\begin{figure*}[htbp]
    \centering
    \begin{minipage}[t]{0.24\textwidth}
        \centering
        \includegraphics[width=\linewidth]{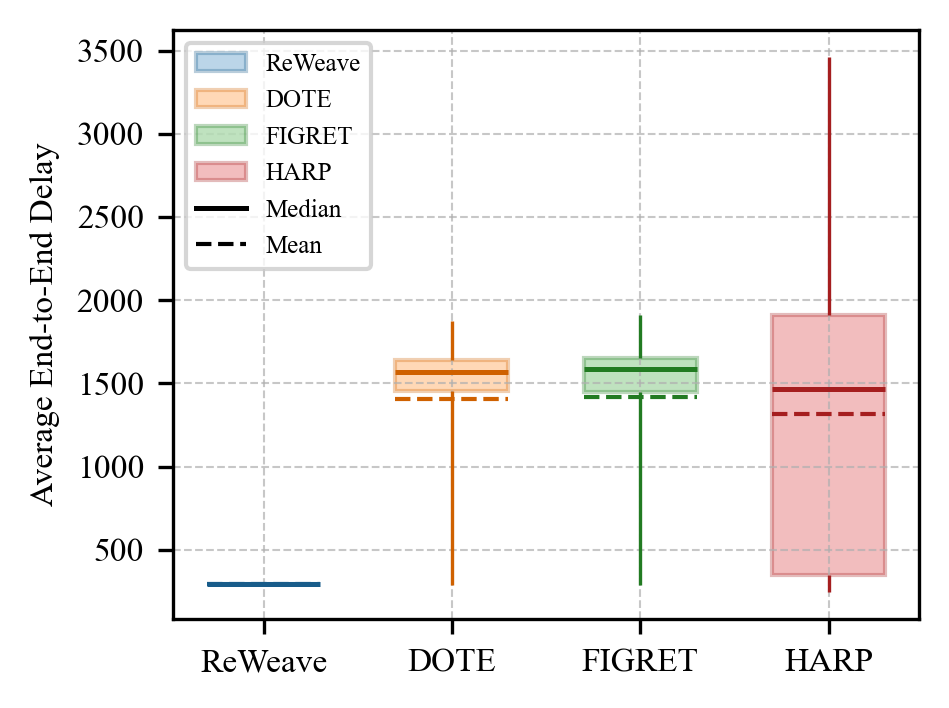}
        \centerline{\small (a) Viatel}
    \end{minipage}
    \hfill
    \begin{minipage}[t]{0.24\textwidth}
        \centering
        \includegraphics[width=\linewidth]{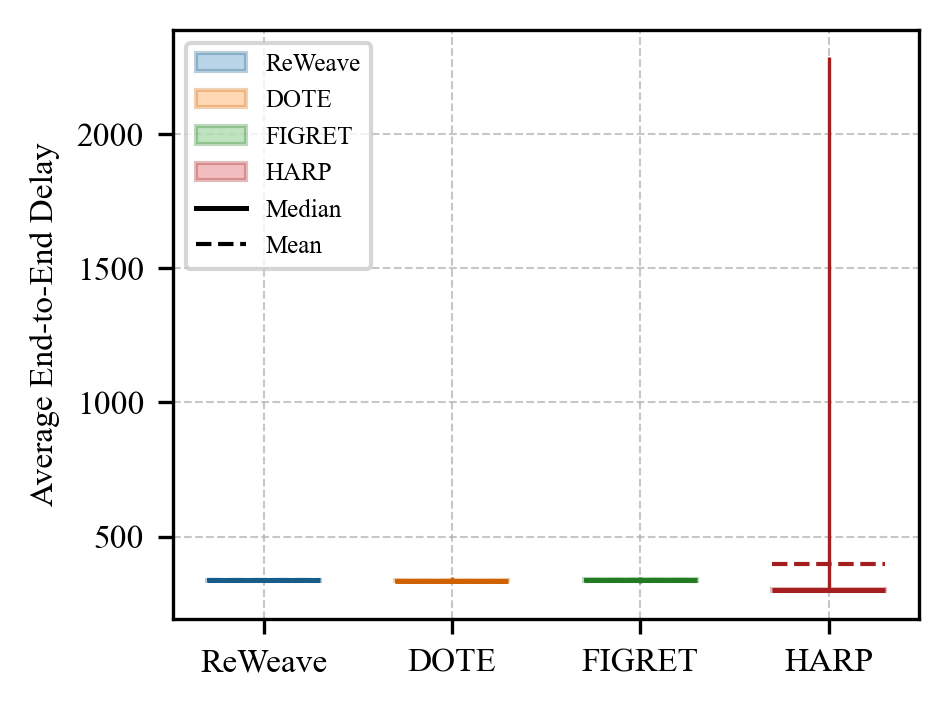}
        \centerline{\small (b) Cogentco}
    \end{minipage}
    \hfill
    \begin{minipage}[t]{0.24\textwidth}
        \centering
        \includegraphics[width=\linewidth]{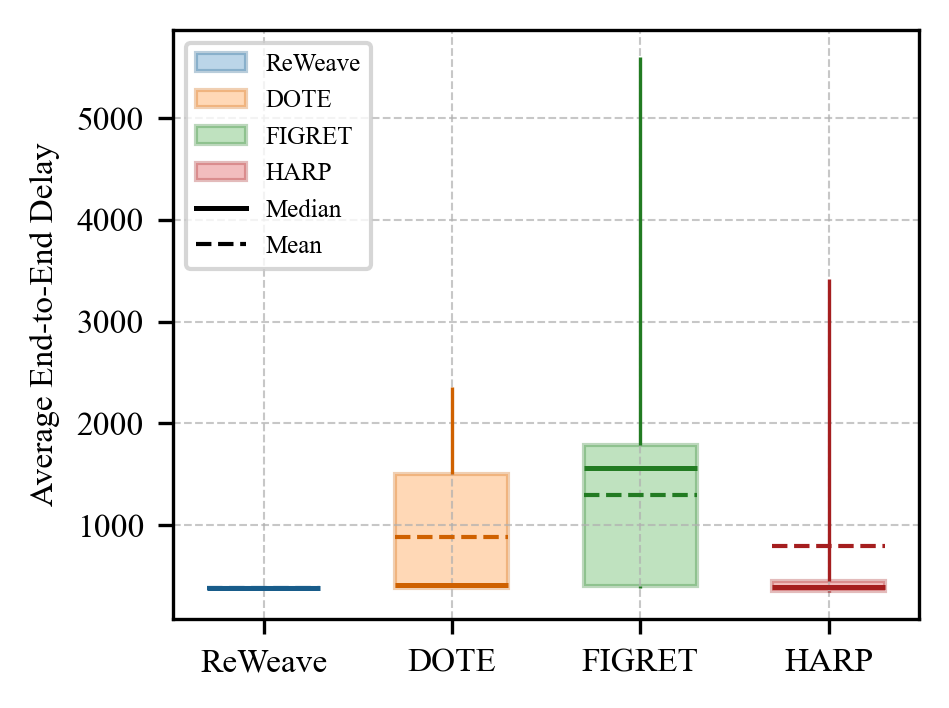}
        \centerline{\small (c) north\_america}
    \end{minipage}
    \hfill
    \begin{minipage}[t]{0.24\textwidth}
        \centering
        \includegraphics[width=\linewidth]{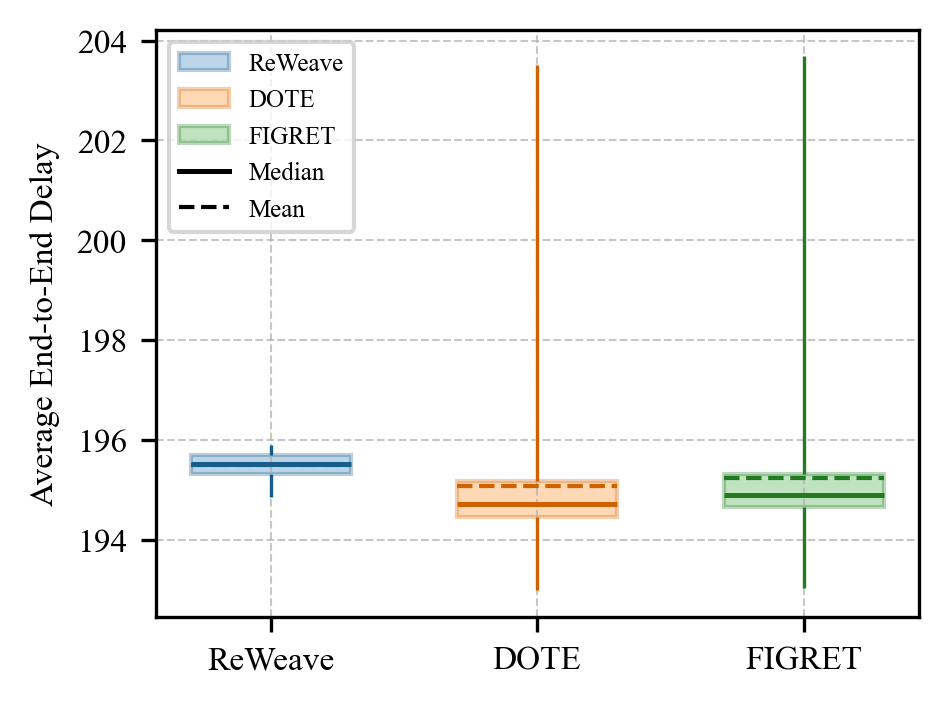}
        \centerline{\small (d) africa}
    \end{minipage}
    \caption{Average end-to-end delay under single-link failure scenarios across different topologies.}
    \label{fig:e2e_delay}
\end{figure*}

We adopt the average end-to-end delay metric as defined in \cite{b16}. Given a traffic matrix and the corresponding TE configuration provided by a scheme, the delay \(\Omega_{TM}\) for a given routing configuration is computed as:
 $\Omega_{TM} = \sum_{e \in E} \frac{l(e)}{C(e) - l(e)}$
where \(C(e)\) denotes the capacity of link \(e\), and \(l(e)\) represents its load under the given TM.
Fig.~\ref{fig:e2e_delay} presents the average end-to-end delay across various TE methods under single-link failure scenarios. Globally, ReWeave demonstrates robust load-balancing behavior by maintaining delay within a controlled range, indicating that link utilization across the topology remains within acceptable limits.
Although other schemes occasionally yield slightly lower delays in certain cases, they suffer from extreme outliers—severe congestion on specific links can dramatically degrade overall network performance.
These findings suggest that the delay introduced by failure-induced rerouting is acceptable and preferable to the substantial performance losses caused by congestion.

\subsection{Routing and Backup Schemes}

We compare different routing methods and backup schemes on the Viatel topology to demonstrate the rationality of our routing approach.
Table~\ref{tab:routing_backup} compares the total training time and the average inference time for each traffic matrix under the KSP and EDKSP routing schemes. 
The EDKSP-based model is significantly faster in both training and inference, as EDKSP's stricter constraints result in fewer paths and a smaller parameter space.
In addition, Table~\ref{tab:routing_backup} summarizes the total number of paths required by each scheme. The model based on EDKSP routing incurs significantly lower overhead compared to the one based on KSP routing. Furthermore, the number of backup paths accounts for only about 6\% of the total paths required for distributing traffic among all node pairs, indicating that backup routing does not impose scalability limitations. This observation is consistent with the design principles of the path selection method described in Section~\ref{sec:IV-B}.

\begin{figure}[htbp]
  \centerline{\includegraphics[width=0.3\textwidth]{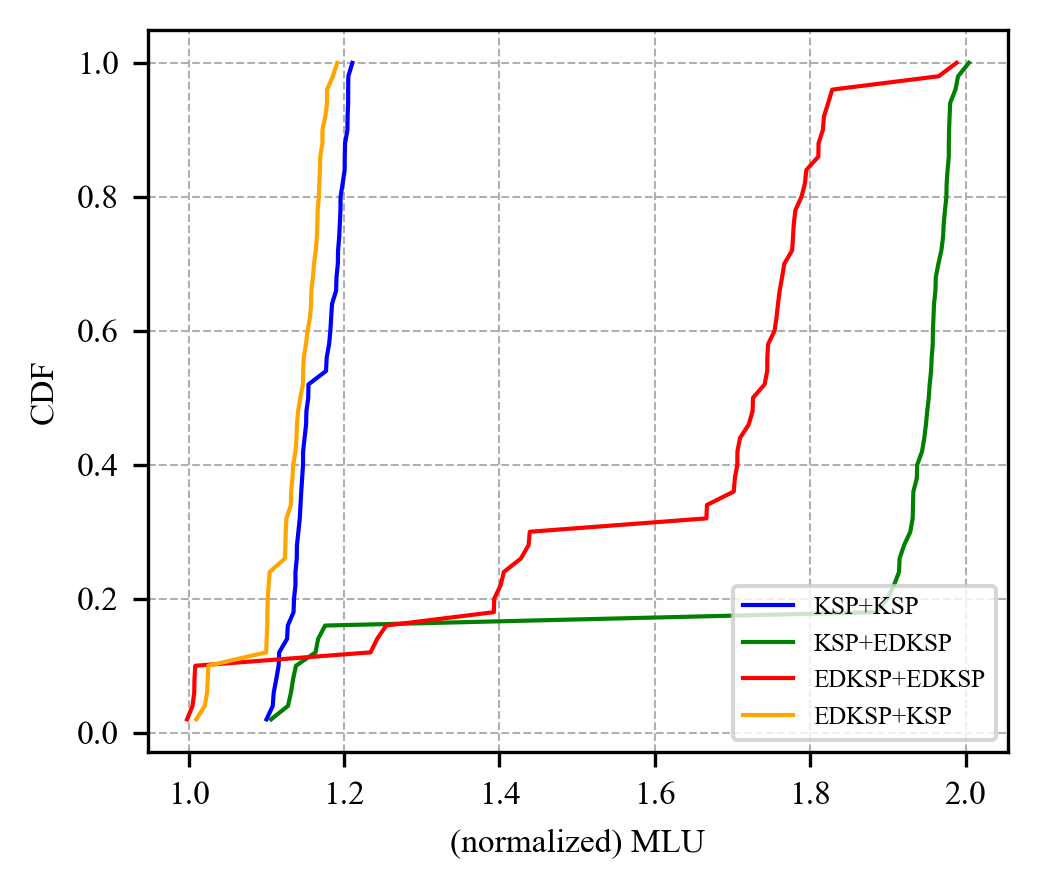}}
  \caption{Comparison of different routing and backup path schemes under single-link failure scenarios.}
  \label{fig:ablation_result}
\end{figure}

Regarding backup paths, KSP as a backup path offers more options than EDKSP, making the KSP-based backup path scheme slightly slower than the EDKSP-based scheme. Nevertheless, this delay is less than 0.001 seconds and thus remains within an acceptable margin. We also examine the performance of different routing and backup combinations under fault scenarios, as shown in Fig.~\ref{fig:ablation_result}. The KSP-based backup path scheme significantly outperforms the EDKSP-based scheme, as KSP provides more options for rerouting traffic upon link failure, while EDKSP often only offers 1-2 paths per node pair, leading to traffic congestion on the same path in case of failure. Overall, using EDKSP as the routing method and KSP as the backup path provides an efficient and balanced tradeoff between performance and speed.

\begin{table}[ht]
    \centering
    \caption{Training \& Inference Time of Routing Schemes}
    \begin{tabular}{|c|c|c|c|}
        \hline
        \textbf{Routing+Backup} & \textbf{Pre. Time (s)} & \textbf{Comp. Time (s)} & \textbf{Total Paths} \\
        \hline
        KSP + KSP & 8588 & 0.07 & 60132 \\
        KSP + EDKSP  & 8588 & 0.06 & 59260 \\ 
        EDKSP + EDKSP   & 1094 & 0.02 & 15463 \\ 
        EDKSP + KSP  & 1094 & 0.02 & 16335 \\ 
        \hline
    \end{tabular}
    \label{tab:routing_backup}
\end{table}

\subsection{Robustness to Unseen Demand Variations}

To evaluate model's sensitivity to traffic demand, we followed the methodology in DOTE\cite{b28} and added random noise to the test matrices. We multiplied each demand by a factor sampled uniformly from $[1 - \alpha, 1 + \alpha]$, for $\alpha \in \{0.1, 0.2, 0.3\}$. Table~\ref{tab:robustness} presents the results. Across all noise levels, change in mean MLU is minimal, and in several cases even shows a slight improvement. Specifically, for the highest noise level ($\alpha=0.3$), we considered the 99th percentile MLU change, which remains well-contained and acceptable. These findings show our model is highly robust to demand uncertainty.

\begin{table}[ht]
  \caption{MLU Performance Change with Traffic Noise}
  \begin{center}
  \begin{tabular}{|l|c|c|c|c|}
    \hline
    \textbf{Topology} 
    & \multicolumn{3}{c|}{\textbf{Mean MLU Change}} 
    & \multicolumn{1}{c|}{\textbf{99th-\%ile Change}} \\
    \hline
    & \textbf{$\alpha=0.1$} & \textbf{$\alpha=0.2$} & \textbf{$\alpha=0.3$} & \textbf{$\alpha=0.3$} \\
    \hline
    Viatel & +0.07\% & +0.16\% & +0.33\% & +2.59\% \\
    Cogentco & -1.21\% & -0.10\% & +3.41\% & -4.33\% \\
    north\_america & -0.04\% & -0.00\% & +0.16\% & +2.11\% \\
    \hline
  \end{tabular}
  \label{tab:robustness}
  \end{center}
\end{table}

\section{Related Work}

Network failures, especially link failures, are common in large-scale deployments\cite{b6}. Cloud providers must maintain performance objectives while handling failures. Given the global scale and rapid evolution of networks, failure-handling mechanisms must be scalable\cite{b2}.

\cite{b11} proposed a unified framework that combines failure recovery with traffic engineering. Link failure handling strategies in TE are generally categorized into protection and restoration. Protection proactively ensures congestion-free operation under a set of predefined failure scenarios. \cite{b11} introduces scenario-specific routing entries at ingress nodes, each with a set of weighted paths. Upon detecting a failure, traffic is rerouted using the preconfigured weights. FFC\cite{b7}, as a powerful hedging-like scheme, achieves congestion-free operation under up to $f$ simultaneous failures via offline optimization and local weighted routing. Extensions such as Teavar and Flexile incorporate the probabilities of failure scenarios, aiming to ensure the network meets performance requirements in a set of scenarios whose total probability is at least  $\beta$. However, these methods tend to be conservative and rely heavily on preset backup paths or scenario-specific weights, which may limit their flexibility.

Early restoration schemes focused on fast local rerouting using weighted splitting to restore connectivity\cite{b8}. These lightweight, demand-agnostic approaches have been integrated into TE methods like DOTE and FIGRET. Notably, the authors of DOTE [25] also suggested incorporating failures into the training process as a potential future direction.However, such methods can sometimes cause congestion. 
Flexile~\cite{b14} reports that failures can cause link loads to exceed capacity by 10\%–20\%, which degrades performance for latency-sensitive applications like web search and video streaming. 
Global rerouting approaches such as HARP and SMORE recompute configurations upon failure. While these global rerouting approaches aim for higher performance by re-optimizing globally, they can introduce response latency due to centralized control, especially in large networks.
It is also possible to consider end-to-end path regeneration when the practical deployment cost is low and paths can be quickly activated.
Emerging SDN technologies provide new opportunities for failure recovery in TE. 
Indeed, local fast rerouting is a mature and well-studied approach for providing rapid, data-plane-level recovery \cite{b30}. 
For instance, Group-Table-based Rerouting (GTR) enables switches and controllers to share equal responsibility\cite{b13}. Only backup paths between adjacent switches need to be computed, avoiding memory overhead from maintaining per-flow backups, and achieving fast responses to single-link failures.


\section{Conclusion}
We presented ReWeave, a robust and scalable traffic engineering scheme designed to handle link failures through localized, SRv6-based path weaving. ReWeave focuses only on the two endpoints of the failed link to perform local detours, enabling rapid failure recovery without compromising load balancing or scalability. We conducted extensive evaluations on a variety of real-world and synthetic network topologies. The results indicate that ReWeave achieves better performance than some existing approaches in terms of reaction speed and link utilization, reducing worst-case link load by up to 40.9\%, while achieving a packet loss rate comparable to offline-optimized solutions. These findings demonstrate the practical potential of ReWeave for large-scale deployment. The principles of ReWeave also open up exciting directions for future work, including comparing our localized method with end-to-end path regeneration, which is possible but not explored in this work, and adapting it for non-SR networks.

\section*{Acknowledgments}
This work was supported by National Key R\&D Program of China under grant No. 2022YFB3102901.
\bibliographystyle{IEEEtran}


\end{document}